\documentclass[useAMS]{mn2e}
\usepackage{times}
\input{epsf}

\title[High-frequency power in black holes]
{High-frequency X-ray variability as a mass estimator of stellar and
supermassive black holes}

\author[M. Gierli{\'n}ski, M. Niko{\l}ajuk and B. Czerny]
{Marek
Gierli{\'n}ski$^{1,2}\thanks{E-mail:Marek.Gierlinski@durham.ac.uk}$,
Marek Niko{\l}ajuk$^{3}$
and Bo{\.z}ena Czerny$^{4}$\\
$^1$Department of Physics, University of Durham, South Road,
Durham DH1 3LE, UK\\
$^2$Astronomical Observatory, Jagiellonian University, Orla 171,
30-244 Krak{\'o}w, Poland\\
$^3$Department of Physics, University of Bia{\l}ystok, Lipowa 41,
15-424 Bia{\l}ystok, Poland\\
$^3$Copernicus Astronomical Centre, Bartycka 18, 00-716 Warszawa, Poland\\
}

\date{Submitted to MNRAS}
\pagerange{\pageref{firstpage}--\pageref{lastpage}} \pubyear{2005}

\begin{document}

\topmargin = -0.5cm

\maketitle

\label{firstpage}

\begin{abstract}

There is increasing evidence that supermassive black holes in active
galactic nuclei (AGN) are scaled-up versions of Galactic black
holes. We show that the amplitude of high-frequency X-ray
variability in the hard spectral state is inversely proportional to
the black hole mass over eight orders of magnitude. We have analyzed
all available hard-state data from {\it RXTE} of seven Galactic
black holes. Their power density spectra change dramatically from
observation to observation, except for the high-frequency ($\ga$ 10
Hz) tail, which seems to have a universal shape, roughly represented
by a power law of index -2. The amplitude of the tail, $C_M$
(extrapolated to 1 Hz), remains approximately constant for a given
source, regardless of the luminosity, unlike the break or QPO
frequencies, which are usually strongly correlated with luminosity.
Comparison with a moderate-luminosity sample of AGN shows that the
amplitude of the tail is a simple function of black hole mass, $C_M
= C/M$, where $C \approx 1.25$ M$_\odot$ Hz$^{-1}$. This makes $C_M$
a robust estimator of the black hole mass which is easy to apply to
low- to moderate-luminosity supermassive black holes. The
high-frequency tail with its universal shape is an invariant feature
of a black hole and, possibly, an imprint of the last stable orbit.

\end{abstract}


\section{Introduction}
\label{sec:introduction}

Astrophysical black holes are very simple objects, completely
characterized by their mass and spin. Hence, the gravitational
potential around a black hole simply scales with its mass. An
important question in high-energy astrophysics is whether the
accretion flow properties scale with the black hole mass in a simple
manner, or, more specifically, whether active galactic nuclei (AGN)
are scaled-up versions of Galactic black hole binaries (BHB).

One of the ways of tackling this problem is to study X-ray
variability, which is observed in accreting black holes of all
masses. Recent advances in mass estimates of AGN central black holes
lead to discovery of dependence of the observed variability
properties on mass. Long X-ray monitoring campaigns allowed to
construct power density spectra (PDS) of accreting supermassive
black holes which turned out to have a roughly of (broken) power-law
shape. The variability amplitude (the excess variance; e.g. Lu \& Yu
2001; Markowitz \& Edelson 2001) and the frequency of the break
(e.g. M$^c$Hardy et al. 2004, 2006) can depend on the black hole
mass.

In order to use the X-ray variability for mass measurement we need a
property which scales only with the black hole mass, and does not
change with accretion rate. The break frequency does not satisfy
this condition, as it changes significantly with the accretion rate,
in X-ray binaries (e.g. Done \& Gierli{\'n}ski 2005). M$^c$Hardy et
al. (2006) showed that a more general relation holds between the
break frequency, $\nu_b$, and the black hole mass, $M$: $\nu_b = A
L_{\rm bol}^B / M^C$, where $A$, $B$ and $C$ are constants. This
relation includes a significant dependence on the source bolometric
luminosity, $L_{\rm bol}$.

It was already suggested by Hayashida et al. (1998) that measuring
the normalization of the high-frequency tail of the power spectrum,
well above the high-frequency break, is an interesting possibility
for black hole mass measurement. Equivalently, one can use the
excess variance, $\sigma^2_{\rm NXS}$, measured for short data sets.
This general line was followed by Czerny et al. (2001), Papadakis
(2004), Niko{\l}ajuk, Papadakis \& Czerny (2004, hereafter N04) and
Niko{\l}ajuk et al. (2006, hereafter N06). However, the method was
not reliably checked against the dependence on the source accretion
rate.

In this paper we put the idea of $\sigma^2_{NXS} \propto M^{-1}$
correlation to the test. We use an extensive set of BHB observations
to see if and when $\sigma^2_{\rm NXS}$ is constant for a given mass
and whether it anticorrelates with the black hole mass.

\section{High-frequency power}
\label{sec:power}

\begin{figure*}
\begin{center}
\leavevmode \epsfxsize=14cm \epsfbox{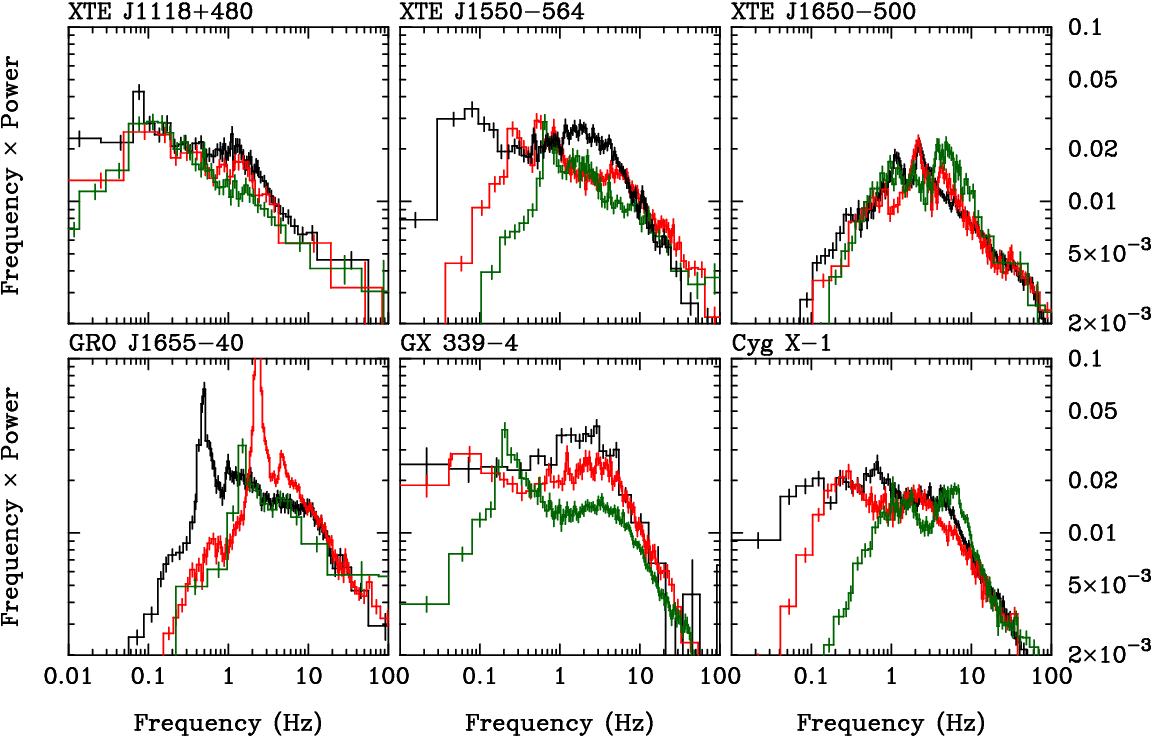}
\end{center}
\caption{Power density spectra of black hole X-ray binaries. Each
panel shows three spectra in the low/hard spectral state, selected
from the data analyzed in Sec. \ref{sec:data}. The selection
demonstrates that despite dramatic changes in the PDS shape at lower
frequencies (including variable QPO), the high-frequency part
($\ga$10 Hz) of the PDS remains relatively constant.}
\label{fig:pds}
\end{figure*}

Power density spectra of many AGN can be approximated by a broken
power law, with power $P_\nu \propto \nu^{-1}$ below and $P_\nu
\propto \nu^{-2}$ above the break frequency, $\nu_b$ (e.g. Markowitz
et al. 2003b), where $P_\nu$ is the power spectral density
normalized to the mean and squared. A second break at lower
frequencies, below which the power is roughly $P_\nu \propto
\nu^{0}$, has been also observed (e.g. Pounds et al. 2001;
Markowitz, Edelson \& Vauhgan 2003a). At the zeroth order of
approximation this is consistent with the PDS observed in
stellar-mass BHB in the hard spectral state. Fig. \ref{fig:pds}
shows a sample of PDS from Galactic BHB in the hard state (details
of the data reduction are described in Section \ref{sec:data}).
Plainly, these spectra are much more complex that a doubly-broken
power law, with multiple broad and narrow noise components, usually
well described by a series of Lorentzians (e.g. Pottschmidt et al.
2003). However, despite this complexity, the entire spectral shape
roughly resembles a (doubly) broken power law.

N04 assumed that the break frequency is inversely proportional to
the black hole mass, while the $P_\nu \propto \nu^{-1}$ part of the
PDS below the break (the `flat top' in $\nu P_\nu$ diagrams) has
constant normalization, independent of the black hole mass. Yet
inspection of several BHB power spectra clearly shows that neither
of these is constant for a given source. The break frequency is
known to change with accretion rate (e.g. Done \& Gierli{\'n}ski
2005). The `flat top' normalization can change as well, as one can
see in GX 339--4 spectra in Fig. \ref{fig:pds}.

There is, however, one feature of these power spectra that remains
remarkably invariant: the high-frequency spectral shape, above
$\nu_b$. For a given source it can be roughly described by a single
power law with constant index of 1.5--2.0, and constant
normalization for various observations differing in luminosity by
more than one order of magnitude. In this paper we test the idea
that the high-frequency part of the PDS remains fairly constant for
a given source and scales with the black hole mass. This is a simple
refinement of the idea proposed by N04 and later developed by N06.

Here we do not make any assumptions about how the characteristic
frequencies (e.g. break frequency) depend on black hole mass.
Instead, we assume that the the PDS {\em above} the break frequency
(the high-frequency tail) has a universal spectral shape (roughly
$\propto \nu^{-2}$) with normalization depending on the black hole
mass. This can be written as \begin{equation}P_\nu = C_M
(\nu/\nu_0)^{-2},\label{eq:tail}\end{equation} where $\nu_0$ is an
arbitrary frequency which we chose to be $\nu_0$ = 1 Hz. Thus, $C_M$
(in units of Hz$^{-1}$) is the normalization of the (extrapolated)
high-frequency tail at 1 Hz.

The assumption that $C_M$ is unique function of the black hole mass
would directly correspond to the original assumption of N04 about
constancy of $P(\nu_b)\nu_b$ if $\nu_b$ were constant for a given
black hole mass. Due to limited statistics it is often difficult to
study details of the high-frequency shape of the PDS. Therefore, we
simplify the situation by calculating the amplitude, or the excess
variance, of variability in a given frequency band significantly
above the break. This can be done directly from a light curve or by
integrating the PDS. The excess variance calculated between
frequencies $\nu_1$ and $\nu_2$ (both greater than $\nu_b$) is:
\begin{equation}\sigma_{\rm NXS}^2 = \int_{\nu_1}^{\nu_2} P_\nu d\nu
= C_M \nu_0 \left( {\nu_0 \over \nu_1} - {\nu_0 \over \nu_2}
\right).\label{eq:sigma_nxs}\end{equation} The key assumption, which
we want to test in this paper, is that $C_M$ is inversely
proportional to the black hole mass, $C_M = C/M$, where $C$ is a
constant. We note that our $C$ is the same constant as constant $C$
defined by N04 in eq. 4, divided by $\nu_0^2$.


\section{Data reduction and selection}
\label{sec:data}

\begin{table}
\begin{tabular}{lcc}
\hline Source Name & Start & End\\
\hline
XTE J1118+480     & 2000-03-29 & 2000-08-08\\
                  & 2005-01-13 & 2005-02-26\\
4U 1543--47       & 2002-06-17 & 2002-10-11\\
XTE J1550--564    & 1998-09-07 & 1999-05-20\\
                  & 2000-04-10 & 2000-07-16\\
                  & 2001-01-28 & 2001-04-29\\
                  & 2002-01-10 & 2002-03-05\\
                  & 2003-03-27 & 2003-05-16\\
XTE J1650--500    & 2001-09-06 & 2002-04-21\\
GRO J1655--40     & 2005-02-20 & 2005-11-11\\
GX 339--4         & 2002-04-02 & 2003-05-06\\
                  & 2003-12-28 & 2005-08-12\\
Cyg X-1           & 1996-02-12 & 2006-01-12\\
\hline
\end{tabular}

\caption{Log of {\it RXTE} observations. Each set of data
corresponds to one transient outburst. For Cyg X-1 we used all data
publicly available in February 2007.}

\label{tab:obslog}
\end{table}


We used publicly available {\it Rossi X-ray Timing Explorer} ({\it
RXTE}) Proportional Counter Array (PCA) data of seven black hole
binaries, listed in Table \ref{tab:obslog}. First, we extracted
background-corrected energy spectra from Standard-2 data (top layer,
detector 2 only) for each pointed observation. These were used to
create hardness-intensity diagrams (HID). Intensity is defined as
the total 2--60 keV count rate and the hardness ratio is the ratio
of count rates in energy bands 6.3--10.5 and 3.8--6.3 keV. We also
calculated power density spectra (PDS) for each observation from
full-band (2--60 keV) data in 0.0039--128 Hz frequency band. We
subtracted the Poissonian noise from the PDS, corrected them for
dead-time effects (Revnivtsev, Gilfanov \& Churazov 2000) and
background (Berger \& van der Klis 1994)

We used HIDs to identify hard state spectra. In transients these
spectra lie on the vertical branch of the diagram at hardness ratio
$\ga$0.8, corresponding to the rise or decay of the outburst. While
in GX 339--4 the transition from the vertical hard branch into
intermediate horizontal branch is abrupt and well defined, in XTE
J1550--564, XTE J1650--500 and GRO J1655--40 we also included a few
points from the part of the HID where the hard branch turns into
intermediate, as they have sufficiently large hardness ratio to be
(potentially) classified as the hard state. In Cyg X-1 we
arbitrarily selected observations with hardness ratio greater than
0.9. We also arbitrarily rejected all data with average count rate
less than 10 counts per s per PCU as they did not have enough
statistics to robustly calculate the high-frequency power. All the
selections are shown in Figs. \ref{fig:cm1} and \ref{fig:cm2}.

For a given source (and separately for each outburst in case of GX
339--4 and XTE J1550--564) we measured the high-frequency power,
$\sigma^2_{{\rm NXS}}$, with their statistical errors, $\alpha$, for
each pointed observation in the hard state. The power (or excess
variance) was measured by integrating the PDS over the frequency
band 10--128 Hz (eq. \ref{eq:sigma_nxs}). Then, we computed the mean
amplitude, $\langle C_M \rangle$ for a given source (or a particular
outburst of the source) weighted by errors of $C_{M}$. The error of
$\langle C_M \rangle$ was estimated from $\chi^2$ statistics for
$\Delta\chi^2 = 2.7$, i.e. corresponding to 90 per cent confidence
limits.

Despite the initial count rate selection, some data sets (in
particular from short observations) gave high-frequency power with
large errors. We discarded all these, setting an arbitrary upper
limit on error, $\alpha < 0.3 \sigma^2_{{\rm NXS}}$.

\section{Results}
\label{sec:results}

\begin{figure}
\begin{center}
\leavevmode \epsfxsize=8.5cm \epsfbox{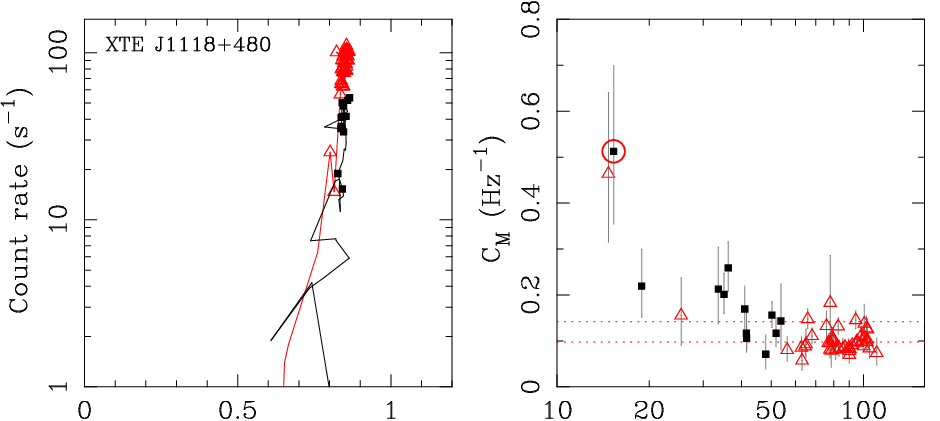} \epsfxsize=8.5cm
\epsfbox{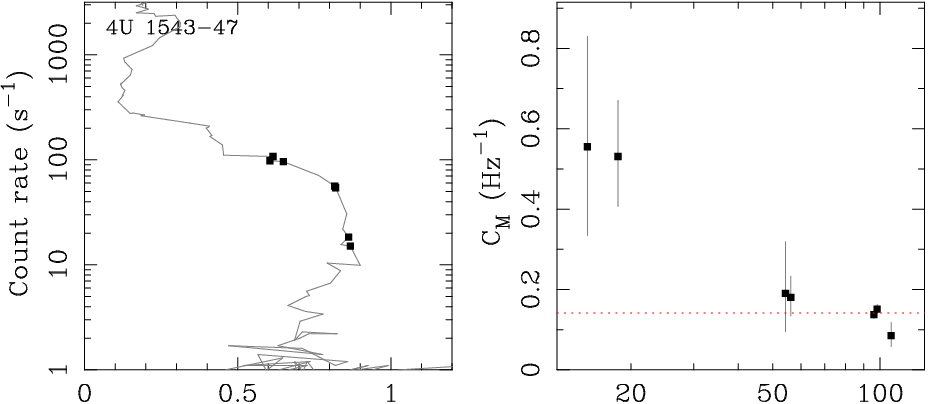} \epsfxsize=8.5cm \epsfbox{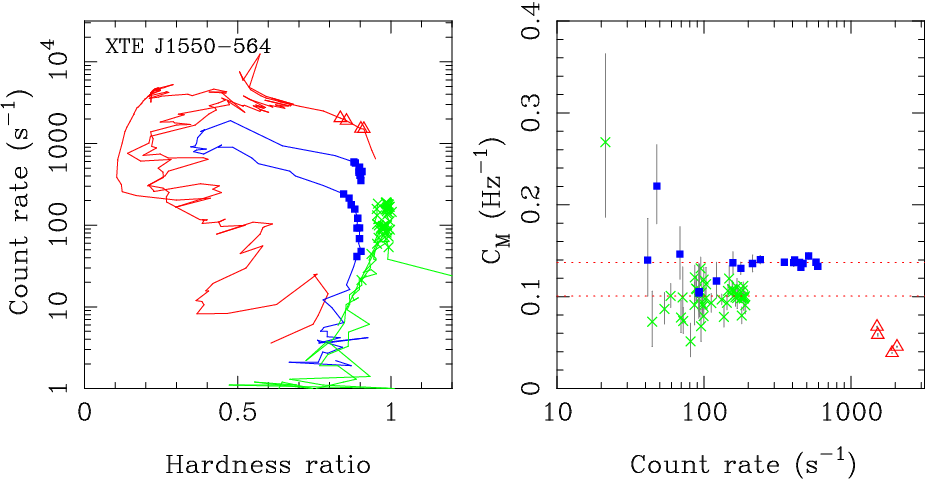} \epsfxsize=8.5cm
\epsfbox{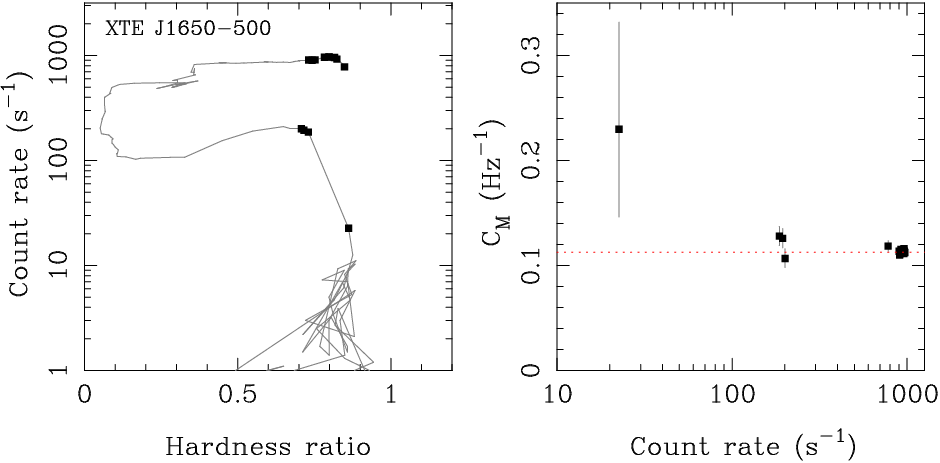}
\end{center}

\caption{Panels on the left show hardness-intensity diagrams. Solid
line shows the entire outburst of a given source, while the data
points correspond to the hard-state selection of pointed
observations, used for further analysis. Diagrams on the right show
the high-frequency variability amplitude, $C_M$ (eq. \ref{eq:tail}),
as a function of count rate, for selected hard-state observations.
The horizontal line shows the weighted average of $C_M$ for the
entire outburst. XTE J1118+480 shows data from two outbursts: 2000
(red triangles) and 2005 (black filled squares). XTE J1550--564
diagram shows data from five outbursts: 1998/1999 (red open
triangles), 2000 (blue squares) and 2001, 2002 and 2003 together
(green crosses). The circled point in XTE J1118+480 panel was used
to estimate background effects (see Section \ref{sec:results}).}
\label{fig:cm1}
\end{figure}

\begin{figure}
\begin{center}
\leavevmode  \epsfxsize=8.5cm \epsfbox{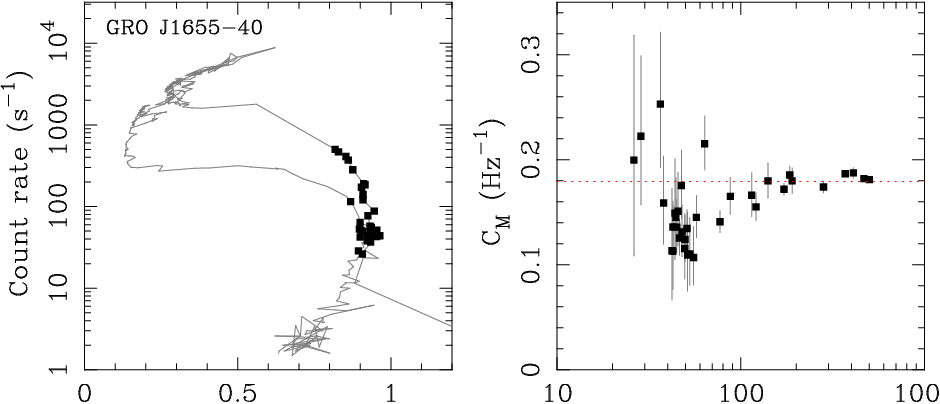} \epsfxsize=8.5cm
\epsfbox{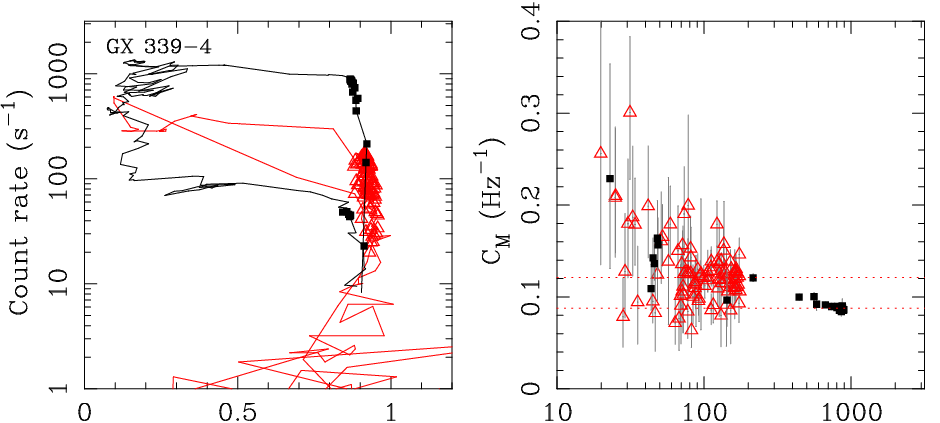} \epsfxsize=8.5cm \epsfbox{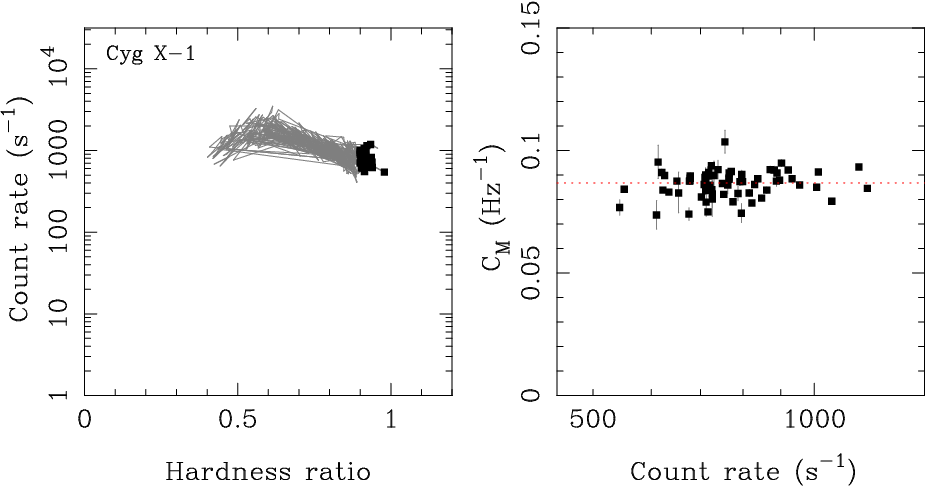}
\end{center}
\caption{Continuation of Fig. \ref{fig:cm1}. GX 339--4 diagrams
shows two outbursts: 2002/2003 (black squares) and 2004 (red
triangles).} \label{fig:cm2}
\end{figure}

Figs. \ref{fig:cm1} and \ref{fig:cm2} show selection of hard-state
data (hardness-intensity diagrams on the left) and measured
amplitude of the high-frequency tail, $C_M$ in the right-hand
panels. It is obvious from these diagrams that $C_M$ is not constant
for a given source and varies from one pointed observation to
another. These variations are not very significant, though. With
very few exceptions (2 observations of GRO J1655--40 and 3 of GX
339--4) individual $C_M$ measurements are within 3$\sigma$ of the
mean. The dispersion is higher in Cyg X-1 where statistics is better
and errors on $C_M$ smaller. Generally, we find that $C_M$ does not
significantly depend on the source brightness (count rate).

\begin{table}
\begin{tabular}{lccc}
\hline
Source Name & Outburst & $\langle C_M \rangle$ (Hz$^{-1}$) & $T_{\rm hard}$ (days)\\
\hline
XTE J1118+480     & 2000 & $0.098\pm0.005$ & 130* \\
                  & 2005 & $0.14\pm0.02$   & 26*  \\
4U 1543--47       & 2002 & $0.14\pm0.01$   & --   \\
XTE J1550--564    & 1998 & $\sim$0.05      & 3    \\
                  & 2000 & $0.137\pm0.002$ & 14   \\
                  & 2001 & $0.095\pm0.001$ & 90*  \\
                  & 2002 & $0.104\pm0.004$ & 50*  \\
                  & 2003 & $0.101\pm0.002$ & 50*  \\
XTE J1650--500    & 2001 & $0.113\pm0.001$ & 8    \\
GRO J1655--40     & 2005 & $0.179\pm0.003$ & 12   \\
GX 339--4         & 2002 & $0.088\pm0.001$ & 20   \\
                  & 2004 & $0.121\pm0.003$ & 180  \\
Cyg X-1 (hard)    &  --  & $0.0868\pm0.0003$ & -- \\
Cyg X-1 (soft)    &  --  & $0.139\pm0.004$   & -- \\
\hline
\end{tabular}

\caption{Mean amplitude of the high-frequency power, $\langle C_M
\rangle$ (eq. \ref{eq:tail}). We show data for each outburst
separately. $T_{\rm hard}$ is approximate duration of the initial
hard state during the rise of the outburst. Asterisks denote
hard-state only outbursts. 4U 1543--47 was not observed in the
initial hard state and Cyg X-1 is a persistent source.}

\label{tab:results}
\end{table}

More significant differences can be found for different outbursts of
a given source. We have analyzed five outbursts of XTE J1550--564.
The three hard-state outbursts between 2001 and 2003 (see Table
\ref{tab:results}) were similar in all properties, so we analyzed
them together. They yield mean $\langle C_M \rangle = 0.101\pm0.002$
Hz$^{-1}$. The hard state in 2000 outburst gave higher $\langle
C_M\rangle = 0.137\pm0.002$ Hz$^{-1}$. However, the 1998 outburst
was very different, with much lower and quickly changing
high-frequency power, $C_M \sim 0.05$ Hz$^{-1}$. We have excluded
the onset of 1998 outburst from further analysis, as it might have
represented a different accretion state (we discuss this in detail
in Sec. \ref{sec:discussion}). Similarly, GX 339--4 and XTE
J1118+480 showed different $C_M$ during different outbursts, though
at least in the case of the latter one this could have been due to
systematic effects at low count rates, which we discuss later in
this section.

\begin{figure}
\begin{center}
\leavevmode \epsfxsize=4cm \epsfbox{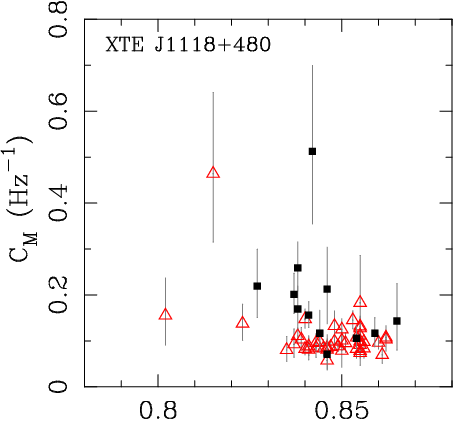} \epsfxsize=4cm \epsfbox{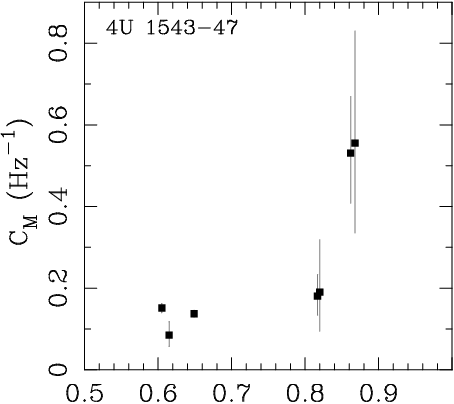}
\epsfxsize=4cm \epsfbox{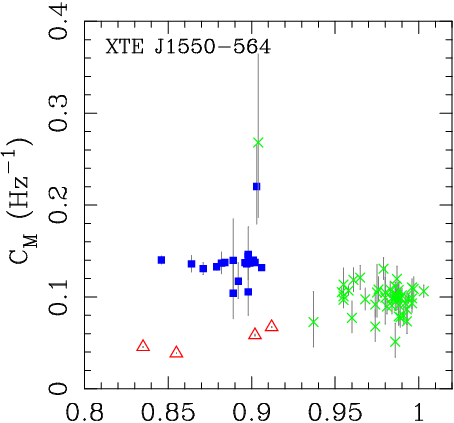} \epsfxsize=4cm \epsfbox{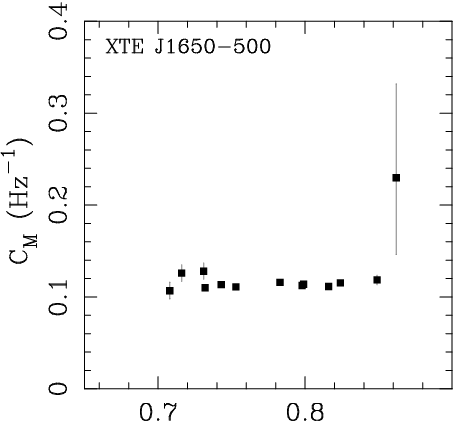}
\epsfxsize=4cm \epsfbox{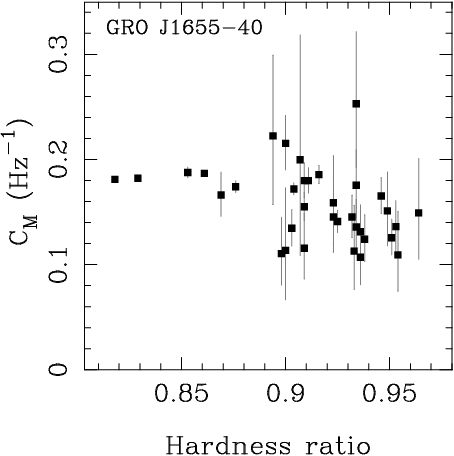} \epsfxsize=4cm \epsfbox{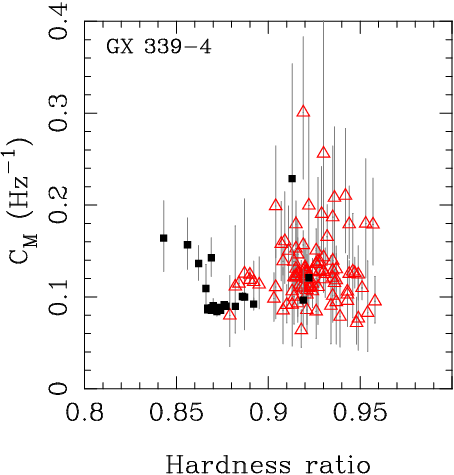}
\end{center}
\caption{Dependence of amplitude of high-frequency variability,
$C_M$, on the hardness ratio. There is no clear trend of $C_M$ as a
function of spectral states.} \label{fig:hrc}
\end{figure}

We looked into the dependence of $C_M$ on the hardness ratio, which
can be regarded as a crude indicator of the spectral state. We did
not find any clear general trend, as illustrated in Fig.
\ref{fig:hrc}.

Additionally, we found variability amplitude for three other black
hole candidates with no black hole mass estimates. The relatively
high value of $\langle C_M \rangle = 0.15_{-0.04}^{+0.03}$ Hz$^{-1}$
obtained for XTE J1720--318 seems to be consistent with a rather low
black hole mass of 5 M$_\odot$, as also suggested by Cadolle Bel
(2004) from disc spectral fitting. For H1743--322 (=XTE J1746--322)
and XTE J1748-288 we found $\langle C_M \rangle = 0.090\pm0.015$ and
$0.056\pm0.008$ Hz$^{-1}$, respectively.

\begin{figure}
\begin{center}
\leavevmode  \epsfxsize=8.5cm \epsfbox{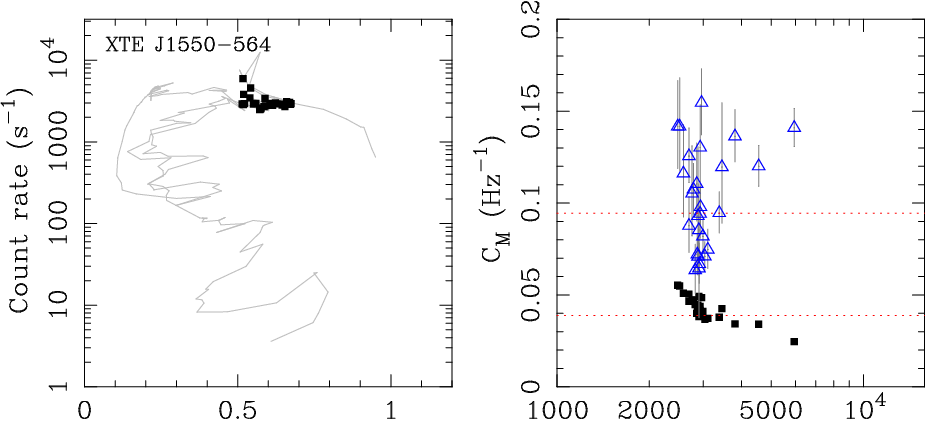} \epsfxsize=8.5cm
\epsfbox{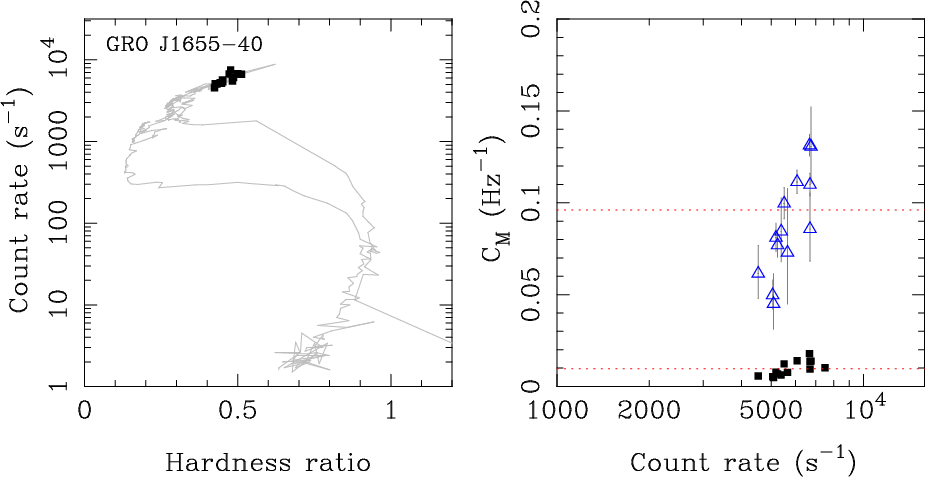}
\end{center}

\caption{Same as in Figs. \ref{fig:cm1} and \ref{fig:cm2}, but for
the very high (steep power law) spectral state. Black filled squares
represent $C_M$ calculated from 2--60 keV PCA data, grey (blue in
colour) open triangles show the high-energy data (roughly 14--60
keV).} \label{fig:vhs}
\end{figure}

\begin{figure}
\begin{center}
\leavevmode  \epsfxsize=8.5cm \epsfbox{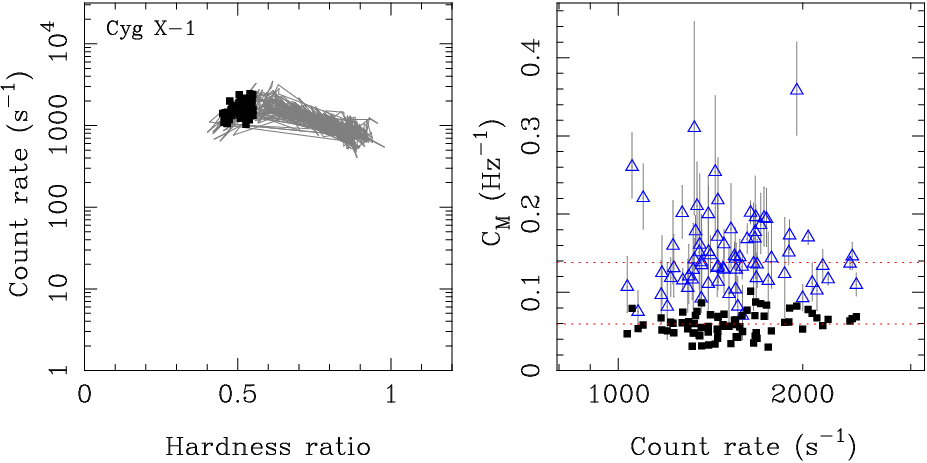}
\end{center}

\caption{Same as in Fig. \ref{fig:vhs}, but for the soft spectral
state of Cyg X-1.} \label{fig:soft}
\end{figure}

The hard X-ray spectral state, which we study in this paper, is
dominated by Comptonized emission. It would be very interesting to
see if similar behaviour can be seen in the other spectral state
dominated by Comptonization, the very high (or steep power law)
state. We have selected the very high state data from
hardness-intensity diagrams of XTE J1550--564 and GRO J1655--40, as
shown in Fig. \ref{fig:vhs}. These observations have significantly
less variability at frequencies $>$10 Hz then the hard state data,
so the resulting $C_M$ is small (black squares in Fig.
\ref{fig:vhs}). But we must remember that the soft X-ray part of the
spectrum (where PCA has highest sensitivity) has strong contribution
from the cold accretion disc, which can dilute the variability
coming from Comptonization (e.g. Done \& Gierli{\'n}ski 2005) hence
suppressing the observed power. Therefore, we extracted additional
power spectra at higher energies (above PCA channel 36, roughly
corresponding to energy of 14 keV), where the disc influence is
negligible. The high-frequency variability amplitude from these data
is much higher, as shown in blue triangles in Fig. \ref{fig:vhs}.
Though there is a large scatter in $C_M$ from individual
observations, the mean $\langle C_M \rangle$ is of order of 0.1
Hz$^{-1}$ in both sources, consistent with the hard state results.
Therefore, the high-frequency variability from Comptonization
appears to be very similar in both hard and very high states (but
see discussion in Sec. \ref{sec:discussion}).

The soft-state data from our BHB sample are strongly dominated by
the disc emission. The only source with reasonable count rates at
higher energies in the soft state is Cyg X-1. We applied the same
approach to Cyg X-1 soft-state data (arbitrarily chosen hardness
ratio in the range 0.45--0.55), as to the very high state data in
the previous paragraph. The result can be seen in Fig.
\ref{fig:soft}. The scatter in individual data points in
significant, with mean $\langle C_M \rangle = 0.138\pm0.004$
Hz$^{-1}$. Since the fit of the constant to the data is rather poor,
$\chi^2_\nu$ = 156/73, the error on $\langle C_M \rangle$ is
indicative only.

\section{Caveats}

We tested our results for possible systematic effects. Many of our
observations have low count rate, which can affect the resulting
power spectra. At low count rates, PCA channels 0--7 can create
artificial power due to problems with the on-board computer
(Revnivtsev, private communication). To test the effect of this we
have removed PCA channels 0--7, where PCA configuration allowed for
that, and calculated new values of $C_M$. We found that this had a
negligible effect on our results.

The white noise level was subtracted from the power spectra during
data reduction. If the white noise level was not estimated
correctly, this could have influenced the amplitude of variability
and the resulting $C_M$. To test this we have selected observations
where high timing resolution was available and calculated power
spectra up to 1024 Hz. As we do not expect any significant power
above a few hundred Hz from black holes (Sunyaev \& Revnivtsev
2000), we assumed that the 512--1024 Hz power could be used as a
good white noise estimator. We reanalyzed these data and calculated
new values of $C_M$. Again, the effect on our results turned out to
be negligible.

Background effects can be potentially important for estimating the
amplitude of variability, which is defined as $({\rm rms} / {\rm
mean})^2$. Our power spectra were calculated from light curves not
corrected for background, so their power is $[{\rm rms} / (R_s +
R_b)]^2$, where $R_s$ and $R_b$ are source and background mean count
rates, respectively. These PDS were then multiplied by $[(R_s + R_b)
/ R_s]^2$, where source and background count rates were estimated
from light curves. In the dimmest observations $R_s$ and $R_b$ are
comparable. To estimate possible background inaccuracy (which is
modelled in the PCA rather then measured) we extracted one PCA and
HEXTE spectrum (using standard HEASARC reduction techniques, adding
1 per cent systematic errors in the PCA) of XTE J1118+480,
corresponding to a high-$C_M$ and low count rate point circled in
Fig. \ref{fig:cm1} (ID 90111-01-02-07, observed on 2005-01-24). We
have fitted the joined PCA/HEXTE spectrum with a simple
Comptonization model (see Done \& Gierli{\'n}ski 2003 for details),
using PCA in 3--40 and HEXTE in 20--200 keV band. The fit was good
($\chi^2 = 137/145$) with no strong residuals. Then we changed the
level of PCA background by $\pm$10 per cent. The fit with 90 per
cent of background was only marginally worse ($\chi^2 = 142/145$),
while the 110 per cent background resulted in a rather poor fit
($\chi^2 = 171/145$). In both cases there were strong residuals in
the PCA and disagreement with HEXTE data above $\sim$25 keV. This
shows that the background in the low count rate data is estimated
with accuracy much better then 10 per cent. Hence, the uncertainty
on $C_M$ due to background estimation is no more than a few per
cent. The increase of $C_M$ by factor 2--5 at low count rates seen
in Fig. \ref{fig:cm1} cannot be caused by incorrect background.
Since we see this effect in most sources below the same count rate
of $\sim$20 s$^{-1}$ (regardless of the distance, hence at different
luminosities) it must be of (unknown) instrumental origin. We would
like to stress, however, that the increase is not statistically
significant, typically less then 2$\sigma$.

The high-frequency variability is known to depend on energy in some
sources (e.g. Nowak et al. 1999). This is important when comparing
BHB with AGN, as we look at different parts of the Comptonized
spectrum, AGN data showing higher scattering orders than BHB. We
have tested our data for energy dependence. This was possible only
in bright observations from XTE J1550-564, GX 339--4 and Cyg X-1,
where statistics at higher energies was good enough. The high-energy
(above $\sim$14 keV) data give $\langle C_M \rangle = 0.13\pm0.01$
Hz$^{-1}$ for the 2000 outburst of XTE J1550--564 and $\langle C_M
\rangle = 0.09\pm0.01$ Hz$^{-1}$ for the 2002 outburst of GX 339--4.
These values are consistent with the broad-band data, which in the
PCA is dominated by soft X-rays (see Table \ref{tab:results}), which
suggests that the high-frequency amplitude is not energy-dependent
in these sources. On the other hand, similar approach to Cyg X-1
gave $\langle C_M \rangle = 0.109\pm0.001$ Hz$^{-1}$, higher by
about 25 per cent higher then the broad-band data.

\begin{figure*}
\begin{center}
\leavevmode \epsfxsize=12cm \epsfbox{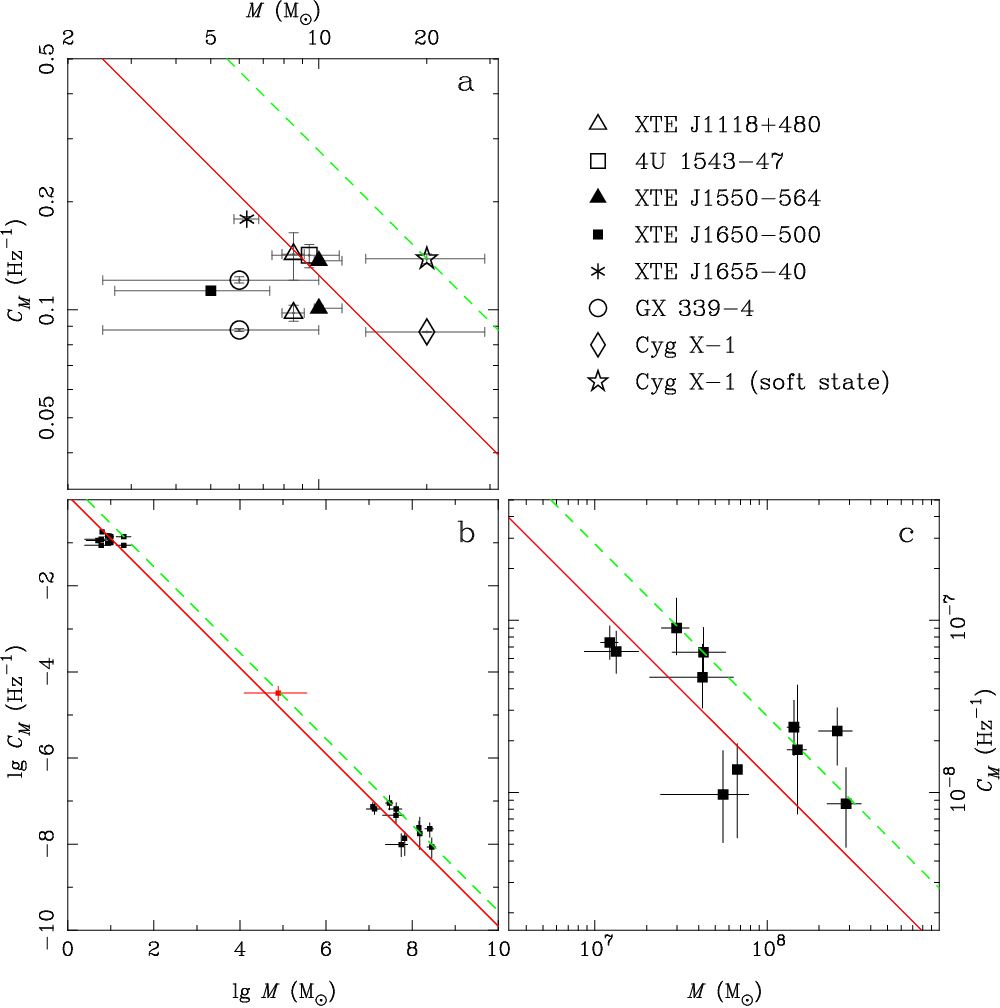}
\end{center}
\caption{Dependence of $C_M$ on black hole mass. Data in the upper
left panel show the results of this paper for X-ray binaries. Black
hole masses, with references, are listed in Table \ref{tab:sources}.
The lower right panel shows the sample of Seyfert 1 galxies from
N06. Black hole masses are obtained from reverberation method
(Peterson et al. 2004). The lower left panel show an overview of
stellar and supermassive black holes. The red cross in the middle of
the diagram represent NGC 4395. The solid red diagonal line shows
the best-fitting relation $C_M = C/M$, with $C = 1.25$ M$_\odot$
Hz$^{-1}$. The green dashed line corresponds to the soft state of
Cyg X-1, with $C = 2.77$ M$_\odot$ Hz$^{-1}$.} \label{fig:mcm}
\end{figure*}


\begin{table}
\begin{tabular}{lcl}
\hline Source Name & $M$ (M$_\odot$) & Reference\\
\hline
XTE J1118+480     & 8.5 (7.9--9.1) & Gelino et al. (2006) \\
4U 1543--47       & 9.4 (7.4--11.4) & Park et al. (2004) \\
XTE J1550--564    & 10 (9.7--11.6) & Orosz et al. (2002) \\
XTE J1650--500    & 5 (2.7--7.3) & Orosz et al. (2004) \\
GRO J1655--40     & 6.3 (5.8--6.8) & Greene, Bailyn \& Orosz (2001)\\
GX 339--4         & 6 (2.5--10) & Cowley et al. (2002) \\
Cyg X-1           & 20 (13.5--29) & Zi{\'o}{\l}kowski (2005) \\
\hline
\end{tabular}

\caption{Black hole masses in X-ray binaries, used in this paper.}

\label{tab:sources}
\end{table}

\section{Dependence on mass}

Fig. \ref{fig:mcm}($a$) shows the dependence of $C_M$ on black holes
mass. We used best currently available mass estimates of black holes
in X-ray binaries, as summarized in Table \ref{tab:sources}. XTE
J1650--500 does not have good mass estimate, though Orosz et al.
(2004) found an upper limit of 7.4 M$_\odot$. We assumed the mass
function, 2.7 M$_\odot$, as the lower limit and we adopted the
actual mass in the middle of this interval, at 5 M$_\odot$. We show
data from different outbursts of XTE 1118+480, XTE J1550--564 and GX
339--4 separately.

Clearly, there is no apparent correlation between black hole mass
and $C_M$, though we must bear in mind that mass estimates in X-ray
binaries are not very accurate. Moreover, different outbursts giving
slightly different $\langle C_M \rangle$ create additional
dispersion. Hence, the overall uncertainties are rather large, so
the expected relation $C_M = C/M$ cannot be robustly confirmed from
X-ray binaries. To do this, we need to extend our studies to
supermassive black holes. N06 compared masses of a sample of Seyfert
1 galaxies measured by reverberation method with masses from
high-frequency variability, using the value of constant $C$ derived
from Cyg X-1 observations. Here we use the same sample of AGN in
order to compare them with our much larger sample of BHB, constrain
the $C_M = C/M$ relation better and get a `big picture' overview of
variability properties for all masses of black holes. In Fig.
\ref{fig:mcm}($c$) we plotted the sample of of N06; panel $b$ shows
the overview of stellar-mass and supermassive black holes.

The red diagonal line in the diagrams represents the best-fitting
function $C_M = C/M$, with $C$ = 1.25$\pm0.06$ M$_\odot$ Hz$^{-1}$.
We would like to point out that this particular value depends on the
selection of X-ray binary data, as different outbursts can give
different $\langle C_M \rangle$. Also, errors on mass are
non-Gaussian in many cases, so the error on $C$, given here, is
indicative only. It is interesting to notice that the constant $C$
is 1.24$\pm$0.06 M$_\odot$ Hz$^{-1}$ for X-ray binaries only and
1.51$_{-0.29}^{+0.37}$ M$_\odot$ Hz$^{-1}$ for AGN only; both values
are consistent within error limits. We also plotted a line (green
dashed) corresponding to the soft state of Cyg X-1, for comparison
(with $C$ = 2.77 M$_\odot$ Hz$^{-1}$). Some of the AGN are
consistent with the soft-state line.

We also tested whether the relation between variability and mass is
really linear. We fitted a more general form of a power-law
dependence, $C_M = C_f (M/$M$_\odot)^{-\alpha}$ to all BHB and AGN
data and found $C_f = 1.19\pm0.05$ Hz$^{-1}$ and index $\alpha =
0.98\pm0.01$ very close to linear relation. Hence, we opt for the
simpler solution and regard the linear relation as well established.

The $C_M = C/M$ relation is in excellent agreement with the data
spanning over eight orders of magnitude in mass. It gives us robust
confirmation of our hypothesis that the high-frequency power is
inversely correlated with black hole mass. Even more firm
confirmation would come from objects with intermediate black holes
mass, filling the big gap in Fig. \ref{fig:mcm}($b$). One category
of sources potentially useful for testing this is the ultra-luminous
X-ray sources (ULX) with possible masses of hundreds of M$_\odot$.
Alas, they are most likely in the very high spectral state, with
soft X-ray emission dominated by the disc. Besides, there are no
reliable high-frequency power spectra available from ULX. An
interesting source for comparison turns out to be a dwarf galaxy NGC
4395, with black hole mass estimates between 0.13 and
3.6$\times10^5$ M$_\odot$ (Kraemer et al. 1999; Ho 2002; Filippenko
\& Ho 2003; Vaughan et al. 2005; Greene \& Ho 2006). The luminosity
is low, $L/L_{\rm Edd} \la 0.1$ (Vaughan et al. 2005). We analyzed
its {\it ASCA} observations of 2000-05-24 and 2000-05-26 and found
the excess variance for this source and $C_M =
3.2_{-1.1}^{+1.4}\times10^{-5}$ Hz$^{-1}$. We show the range of
masses and $C_M$ for NGC 4395 in Fig. \ref{fig:mcm}($b$). They are
in excellent agreement with our mass-variability relation! We can
also give an independent estimate of the black hole mass in NGC 4395
based on the best-fitting value of $C$: $M \approx 3.9\times10^4$
M$_\odot$.

\section{Discussion}
\label{sec:discussion}

We have analyzed all available hard-state data from seven Galactic
black hole systems and found that the amplitude of the
high-frequency tail, $C_M$, is roughly constant for a given source,
changing by no more than a factor two. There is no apparent
dependence of $C_M$ on luminosity or hardness ratio. This contrasts
QPO or break frequency behaviour, which typically show a strong
correlation with luminosity. The scale of change of $C_M$ is also
much less than the observed span of QPO/break frequency, which can
be one order of magnitude within the hard state. This makes $C_M$
much more invariant feature of a given black hole and a robust
estimator of its mass.

There are, certainly, some departures from this rule. Different
outbursts of the same transient can have slightly different
high-frequency tail amplitudes. Most of the sources show an increase
in $C_M$ below count rate of $\sim$20 s$^{-1}$ per PCU. This might
be attributed to systematic instrumental effects at low count rates.
Due to large errors this has a negligible effect on our results.
But, if this effect extends to higher count rates, than the
difference between 2000 and 2005 outbursts of XTE J1118+480
(increase of $C_M$ below 40 s$^{-1}$, see Fig. \ref{fig:cm1}) could
be of instrumental origin.

XTE J1550--564 gives a much clearer exception to this rule. The
first outburst in 1998 had a very short initial hard state with the
high-frequency tail amplitude much lower than in the next outbursts.
In contrast, the three hard-state only outbursts in 2001, 2002 and
2003 produced very consistent results. The explanation of this
phenomenon might lie in the stability of the accretion flow. The
spectral state during the onset of the 1998 outburst was changing
faster than in any other outburst analyzed here and took a different
track on the colour-colour diagram (Done \& Gierli{\'n}ski 2003). It
is possible that the accretion flow during such a rapid transition
was in a different state than in slower transitions, possibly due to
higher ionization state of the reflector (Wilson \& Done 2001; Done
\& Gierli{\'n}ski 2003). Clearly, this also affected its variability
properties, decreasing the high-frequency power. This particular
hard state was different in many aspects, so we rejected it from our
sample. The more stable hard-state only outbursts give probably a
better estimate of $C_M$. Similarly, the persistent source, Cyg X-1,
gave a very stable and robust high-frequency power.

Thus, fast transitions, characterized by an unstable and quickly
changing accretion flow, can apparently break the $C_M = C/M$ law.
In Table \ref{tab:results} we show the approximate duration of the
initial hard state after the onset of the outburst. XTE J1650--500
showed a rather quick transition, so might have suffered from a
similar instability, as XTE J1550--564 in 1998. On the other hand,
the tail amplitude was very stable and returned to the same level in
the hard state at the end of the outburst (see Fig. \ref{fig:cm1}).
The 2004/2005 outburst of GX 339--4 had a much longer hard state
than the 2002/2003 outburst, so perhaps it was a better estimate of
the high-frequency power.

In this work we assumed the constant power-law high-frequency tail
with the spectral index of -2. This is not necessarily true and
detailed fits to PDS of X-ray binaries, where statistics is
sufficient, show indices between 1.5 and 2.0. We note that the
particular choice of the spectral index does not affect the overall
result of our paper, as different index would only introduced a
constant offset in $C_M$. Scatter in spectral indices from one
observation to another might introduce some additional uncertainty,
though this would be very difficult to estimate for AGN.

The luminosities of the Seyfert 1 sample used in this paper are
generally low, $\la$5 per cent of $L_{\rm Edd}$, except for 3C120
and NGC 7469, which are significantly brighter. The 3--10 keV photon
power-law indices are $\Gamma \sim$ 1.5--1.8 (Nandra et al. 1997).
This suggests that these sources are in the hard X-ray spectral
state. On the other hand, the spectral index of Comptonized
component alone is not enough to establish the X-ray spectral state.
Many BHB show soft-state disc-dominated spectra with a flat ($\Gamma
\approx 2$) power-law Comptonized tail, while measuring the {\em
intrinsic} spectral index in AGN is not straightforward due to
presence of complex absorption and/or reflection (Gierli{\'n}ski \&
Done 2004). We cannot rule out (in particular for brighter AGN) that
some of the sources in the sample are actually in the soft X-ray
spectral state. Despite that the mass-variability amplitude
correlation seems to hold well for all of them. Some of the objects
are more consistent with the soft-state Cyg X-1 line in Fig.
\ref{fig:mcm}($c$). On the other hand, the brightest 3C120 and NGC
7469 lay {\em below} the hard-state line. Clearly, the dependence of
high-frequency amplitude on luminosity and spectral state in AGN
requires further studies.

Our analysis of the very high (steep power law) state shows that
after a simple bandwidth correction the tail power is consistent
with the hard-state results. This is very encouraging, showing that
perhaps the universal shape of the high-frequency tail is the
inherent property of Comptonization, regardless of the spectral
state. Narrow-line Seyfert 1 galaxies (NLS1) are most likely the
supermassive counterparts of Galactic sources in the bright very
high state (e.g. Pounds, Done \& Osborne 1995). N04, and later
Niko{\l}ajuk, Gurynowicz \& Czerny (2007) showed that the excess
variance from the NLS1 sample is by a factor $\sim$20 larger than
expected from $C_M = C/M$ correlation established for
moderate-luminosity AGN. NLS1 have also systematically higher break
frequencies for a given mass, so the luminosity dependence of the
break frequency (M$^c$Hardy et al. 2006) can perhaps be related to
the increase in $C_M$. One explanation could be a bandpass effect.
The seed photons in X-ray binaries in the very high state are at
$\sim$1 keV, while in AGN they are at much lower energies, $\la$10
eV. Therefore, we see a different part of the Comptonized spectrum
in X-rays. Gierli{\'n}ski \& Zdziarski (2005) showed that the
variability amplitude strongly increases with energy in the very
high state, while it is almost constant in the hard state. As we see
higher orders of scattering in NLS1 than in the very-high-state
stellar mass black holes, we expect higher variability in the former
ones, as observed. Another possible explanation is contribution from
complex absorption to the observed variability, which can
significantly increase the rms at energies $\sim$1--2 keV (Markowitz
et al. 2003a; Gierli{\'n}ski \& Done 2006). We expect more complex
absorption from bright NLS1 sources with strong outflows. The low-
and moderate-luminosity AGN are less likely to be affected by this
additional variability.

Fig. \ref{fig:mcm} shows that the amplitude  of the high-frequency
tail in the PDS scales very well with the black hole mass. The
correlation holds for over seven orders of magnitude. Thus, the
high-frequency power in accreting black holes in the hard spectral
states seems to be universal. This part of the PDS corresponds to
the shortest timescales producing power in the accretion flow.
Features (e.g. QPOs) occasionally observed at higher frequencies
(e.g. Strohmayer 2001) have much less power. We do not understand
the origin of rapid X-ray variability from accretion flows very
well. It might be created by fluctuations propagating in the
accretion flow (Lyubarskii 1997). The closer to the centre, the
shorter characteristic timescales of fluctuations. Inevitable, there
is a final barrier for propagation, the last stable orbit, which
acts as a low-pass filter, removing all frequencies higher than a
certain limit. The limiting frequency is inversely proportional to
the radius of the last stable orbit and hence to the black hole
mass. With a certain shape of the filter any initial spectrum of
fluctuations will be truncated to a similar shape at higher
frequencies. This might (at least qualitatively) explain the
observed universal shape of the high-frequency tail (Done,
Gierli{\'n}ski \& Kubota 2007).

Obviously, the size of the last stable orbit depends not only on the
black hole mass, but also on its spin. Higher spin would shrink the
last stable orbit as if the black hole mass was smaller. This would
increase $C_M$ by factor 4.8 for a maximally spinning Kerr black
hole with respect to a Schwarzschild one. Alas, the black hole spin
is notoriously difficult to measure (compare, e.g., McClintock et
al. 2006 and Middleton, Done \& Gierli{\'n}ski 2006). The
potentially highly spinning GRS 1915+105 has never been observed in
the low-luminosity hard spectral state. Another culprit is XTE
J1650--500, in which a broad iron line, suggesting high black hole
spin, has been reported (Miller et al. 2002, but see also Done \&
Gierli{\'n}ski 2006). However, as one can see in Fig. \ref{fig:mcm}
$C_M$ of this source is well below the $C/M$ line, so its high spin
does not seem to be supported by our data. On the other hand the
mass of black hole in this source is not very well established, so
we cannot make any strong statements about it. Fig. \ref{fig:mcm}
shows that the scatter in $C_M$ for all sources is only factor two,
so we do not expect large scatter in black hole spins in the sample.

Yet another factor that might influence the results is the
inclination of the disc with respect to the observer. If rapid X-ray
variability is produced in flares or fluctuations corotating with
the disc, then it is affected by Doppler effects for highly inclined
discs. {\.Z}ycki \& Nied{\'z}wiecki (2005) calculated these effects
and predicted that high inclinations would give rise to strong
increase in the high-frequency power. However, the additional signal
appears above $\sim$100 Hz and would not be easily detected in PDS.
We calculated $C_M$ from 10--128 Hz band, so our results should not
be affected by the inclination.

\section{Comparison with break-frequency scaling}
\label{sec:comparison}

An alternative approach of linking variability with black hole mass
is the mass-luminosity-break frequency scaling, $\nu_b = A L_{\rm
bol}^B / M^C$ (M$^c$Hardy et al. 2006; K{\"o}rding et al. 2007),
where the three observables: mass, luminosity and break frequency
form a `fundamental plane' along which the BHB and AGN data
correlate.

We would like to point out few advantages of the method proposed in
this paper over the break-frequency scaling. Generally, it is easier
to find variability amplitude than the break frequency, as the
latter requires modelling of the power spectrum (but see Pessah
2007). The amplitude method does not involve luminosity or accretion
rate, so it doesn't require distance to the source, which in case of
many Galactic black holes is poorly constrained. Another advantage
of this approach is its simplicity, as it relates amplitude of
variability with black hole mass directly, with just one scaling
constant. The break-frequency method requires three independent
constants. This suggests that high-frequency variability amplitude
is more fundamental in nature.

One of the drawbacks of the amplitude method is its limitation to
the hard X-ray spectral state in BHB. Another state dominated by
Comptonization, the very high (steep power law) state is potentially
useful, but its application to bright AGN requires better
understanding of energy dependence of rms. Soft-state spectra of BHB
are strongly diluted by the (stable) disc and good-statistics
high-energy data, required to establish $C_M$ reliably, is not
available (except for Cyg X-1 and, perhaps, GRS 1915+105). But most
of the AGN spectra in the 2--10 keV band are dominated by
Comptonization, so this method might be valid in the soft state. The
same problem seems to affect the break-frequency scaling.
K{\"o}rding et al. (2007) point out that their method is mostly
limited to the hard state, as measuring and defining the break
frequency in soft and very high states is very difficult.

Both break-frequency and amplitude correlations require a shift in
the relation when applied to Cyg X-1 in the soft spectral state
(K{\"o}rding et al. 2007), though this is rather difficult to extend
to soft states of other BHB.

The high-frequency amplitude depends, to some extend, on energy,
while the break-frequency is energy-independent. Another potential
disadvantage of the amplitude method is additional variability
introduced by ionized smeared absorption or reflection in some
sources, which is most pronounced around 1--2 keV (Markowitz et al.
2003b; Gierli{\'n}ski \& Done 2006; Crummy et al. 2006).

\section{Conclusions}
\label{sec:conclusions}

Black hole X-ray binaries have a universal shape of the
high-frequency tail (above the break frequency) in their PDS, as
illustrated in Fig. \ref{fig:pds}. Though the exact shape of the
tail is not easy to establish, it can be approximated by a power
law, $P_\nu = C_M(\nu/\nu_0)^{-2}$. The amplitude, $C_M$ of the tail
is remarkably constant for any given BHB in the hard state,
regardless of the luminosity. When extended to supermassive black
holes in moderate luminosity AGN, the tail amplitude scales very
well with black hole mass, $C_M = C/M$. The best-fitting value of
the scaling constant from our sample of BHB and AGN is $C$ =
1.25$\pm0.06$ M$_\odot$ Hz$^{-1}$. This method can be applied to
estimate black hole masses in many AGN. If the universal shape of
the high-frequency tail is an inherent property of Comptonization,
it might be applied to other spectral states. We speculate that the
constancy of the tail is an imprint of the last stable orbit around
the black hole.

\section*{Acknowledgements}

We thank the anonymous referee for their valuable comments. MG
acknowledges support through a PPARC PDRF and Polish Ministry of
Science and Higher Education grant 1P03D08127. MN and BC
acknowledges support through Polish Ministry of Science and Higher
Education grant 1P03D00829.


\label{lastpage}


\begin{thebibliography}{99}

\bibitem[\protect\citeauthoryear{Berger \& van der
Klis}{1994}]{1994A&A...292..175B} Berger M., van der Klis M., 1994,
A\&A, 292, 175

\bibitem[\protect\citeauthoryear{Cadolle Bel et
al.}{2004}]{2004A&A...426..659C} Cadolle Bel M., et al., 2004, A\&A,
426, 659

\bibitem[\protect\citeauthoryear{Cowley et al.}{2002}]{2002AJ....123.1741C}
Cowley A.~P., Schmidtke P.~C., Hutchings J.~B., Crampton D., 2002,
AJ, 123, 1741

\bibitem[\protect\citeauthoryear{Crummy et al.}{2006}]{2006MNRAS.365.1067C}
Crummy J., Fabian A.~C., Gallo L., Ross R.~R., 2006, MNRAS, 365,
1067

\bibitem[\protect\citeauthoryear{Czerny et al.}{2001}]{2001MNRAS.325..865C}
Czerny B., Niko{\l}ajuk M., Piasecki M., Kuraszkiewicz J., 2001,
MNRAS, 325, 865

\bibitem[\protect\citeauthoryear{Done \& Gierli{\'n}ski}{2003}]{2003MNRAS.342.1041D}
Done C., Gierli{\'n}ski M., 2003, MNRAS, 342, 1041

\bibitem[\protect\citeauthoryear{Done \&
Gierli{\'n}ski}{2005}]{2005MNRAS.364..208D} Done C., Gierli{\'n}ski
M., 2005, MNRAS, 364, 208

\bibitem[\protect\citeauthoryear{Done \&
Gierli{\'n}ski}{2006}]{2006MNRAS.367..659D} Done C., Gierli{\'n}ski
M., 2006, MNRAS, 367, 659

\bibitem[\protect\citeauthoryear{Done, Gierlinski, \&
Kubota}{2007}]{2007arXiv0708.0148D} Done C., Gierli{\'n}ski M.,
Kubota A., 2007, A\&AR, in press, arXiv:0708.0148

\bibitem[Filippenko \& Ho(2003)]{2003ApJ...588L..13F} Filippenko, A.~V., \&
Ho, L.~C.\ 2003, ApJ, 588, L13

\bibitem[\protect\citeauthoryear{Gelino et al.}{2006}]{2006ApJ...642..438G}
Gelino D.~M., Balman {\c S}., K{\i}z{\i}lo{\u g}lu {\"U}., Y{\i}lmaz
A., Kalemci E., Tomsick J.~A., 2006, ApJ, 642, 438

\bibitem[\protect\citeauthoryear{Gierli{\'n}ski \&
Done}{2004}]{2004MNRAS.349L...7G} Gierli{\'n}ski M., Done C., 2004,
MNRAS, 349, L7

\bibitem[\protect\citeauthoryear{Gierli{\'n}ski \&
Zdziarski}{2005}]{2005MNRAS.363.1349G} Gierli{\'n}ski M., Zdziarski
A.~A., 2005, MNRAS, 363, 1349

\bibitem[\protect\citeauthoryear{Gierli{\'n}ski \&
Done}{2006}]{2006MNRAS.371L..16G} Gierli{\'n}ski M., Done C., 2006,
MNRAS, 371, L16

\bibitem[Greene \& Ho(2006)]{2006ApJ...641L..21G} Greene, J.~E., \& Ho,
L.~C.\ 2006, ApJ, 641, L21

\bibitem[\protect\citeauthoryear{Greene, Bailyn, \&
Orosz}{2001}]{2001ApJ...554.1290G} Greene J., Bailyn C.~D., Orosz
J.~A., 2001, ApJ, 554, 1290

\bibitem[\protect\citeauthoryear{Hayashida et
al.}{1998}]{1998ApJ...500..642H} Hayashida K., Miyamoto S., Kitamoto
S., Negoro H., Inoue H., 1998, ApJ, 500, 642

\bibitem[Ho(2002)]{2002ApJ...564..120H} Ho, L.~C.\ 2002, ApJ, 564, 120

\bibitem[\protect\citeauthoryear{K{\"o}rding et
al.}{2007}]{2007MNRAS.380..301K} K{\"o}rding E.~G., Migliari S.,
Fender R., Belloni T., Knigge C., M$^c$Hardy I., 2007, MNRAS, 380,
301

\bibitem[Kraemer et al.(1999)]{1999ApJ...520..564K} Kraemer, S.~B., Ho,
L.~C., Crenshaw, D.~M., Shields, J.~C., \& Filippenko, A.~V.\ 1999,
ApJ, 520, 564

\bibitem[\protect\citeauthoryear{Lu \& Yu}{2001}]{2001MNRAS.324..653L} Lu
Y., Yu Q., 2001, MNRAS, 324, 653

\bibitem[\protect\citeauthoryear{Lyubarskii}{1997}]{1997MNRAS.292..679L}
Lyubarskii Y.~E., 1997, MNRAS, 292, 679

\bibitem[\protect\citeauthoryear{Markowitz \&
Edelson}{2001}]{2001ApJ...547..684M} Markowitz A., Edelson R., 2001,
ApJ, 547, 684

\bibitem[\protect\citeauthoryear{Markowitz \&
Uttley}{2005}]{2005ApJ...625L..39M} Markowitz A., Uttley P., 2005,
ApJ, 625, L39

\bibitem[\protect\citeauthoryear{Markowitz, Edelson, \&
Vaughan}{2003}]{2003ApJ...598..935M} Markowitz A., Edelson R.,
Vaughan S., 2003a, ApJ, 598, 935 

\bibitem[\protect\citeauthoryear{Markowitz et
al.}{2003}]{2003ApJ...593...96M} Markowitz A., et al., 2003b, ApJ,
593, 96 

\bibitem[\protect\citeauthoryear{McClintock et
al.}{2006}]{2006ApJ...652..518M} McClintock J.~E., Shafee R.,
Narayan R., Remillard R.~A., Davis S.~W., Li L.-X., 2006, ApJ, 652,
518

\bibitem[\protect\citeauthoryear{McHardy et
al.}{2004}]{2004MNRAS.348..783M} M$^c$Hardy I.~M., Papadakis I.~E.,
Uttley P., Page M.~J., Mason K.~O., 2004, MNRAS, 348, 783

\bibitem[\protect\citeauthoryear{McHardy et
al.}{2006}]{2006Natur.444..730M} M$^c$Hardy I.~M., Koerding E.,
Knigge C., Uttley P., Fender R.~P., 2006, Natur, 444, 730

\bibitem[\protect\citeauthoryear{Middleton et
al.}{2006}]{2006MNRAS.373.1004M} Middleton M., Done C.,
Gierli{\'n}ski M., Davis S.~W., 2006, MNRAS, 373, 1004

\bibitem[\protect\citeauthoryear{Miller et al.}{2002}]{2002ApJ...570L..69M}
Miller J.~M., et al., 2002, ApJ, 570, L69

\bibitem[\protect\citeauthoryear{Nandra et al.}{1997}]{1997ApJ...477..602N}
Nandra K., George I.~M., Mushotzky R.~F., Turner T.~J., Yaqoob T.,
1997, ApJ, 477, 602

\bibitem[\protect\citeauthoryear{Niko{\l}ajuk, Papadakis, \&
Czerny}{2004}]{2004MNRAS.350L..26N} Niko{\l}ajuk M., Papadakis
I.~E., Czerny B., 2004, MNRAS, 350, L26 (N04)

\bibitem[]{}Niko{\l}ajuk M., Gurynowicz P., Czerny B., 2007,
preprint, astro-ph/0612376

\bibitem[\protect\citeauthoryear{Niko{\l}ajuk et
al.}{2006}]{2006MNRAS.370.1534N} Niko{\l}ajuk M., Czerny B.,
Zi{\'o}{\l}kowski J., Gierli{\'n}ski M., 2006, MNRAS, 370, 1534
(N06)

\bibitem[\protect\citeauthoryear{Nowak et al.}{1999}]{1999ApJ...510..874N}
Nowak M.~A., Vaughan B.~A., Wilms J., Dove J.~B., Begelman M.~C.,
1999, ApJ, 510, 874

\bibitem[\protect\citeauthoryear{Orosz et al.}{2002}]{2002ApJ...568..845O}
Orosz J.~A., et al., 2002, ApJ, 568, 845

\bibitem[\protect\citeauthoryear{Orosz et al.}{2004}]{2004ApJ...616..376O}
Orosz J.~A., McClintock J.~E., Remillard R.~A., Corbel S., 2004,
ApJ, 616, 376

\bibitem[\protect\citeauthoryear{Papadakis}{2004}]{2004MNRAS.348..207P}
Papadakis I.~E., 2004, MNRAS, 348, 207

\bibitem[\protect\citeauthoryear{Park et al.}{2004}]{2004ApJ...610..378P}
Park S.~Q., et al., 2004, ApJ, 610, 378

\bibitem[\protect\citeauthoryear{Pessah}{2007}]{2007ApJ...655...66P} Pessah
M.~E., 2007, ApJ, 655, 66

\bibitem[Peterson et al.(2005)]{2005ApJ...632..799P} Peterson, B.~M., et
al.\ 2005, ApJ, 632, 799

\bibitem[\protect\citeauthoryear{Pottschmidt et
al.}{2003}]{2003A&A...407.1039P} Pottschmidt K., et al., 2003, A\&A,
407, 1039

\bibitem[\protect\citeauthoryear{Pounds, Done, \&
Osborne}{1995}]{1995MNRAS.277L...5P} Pounds K.~A., Done C., Osborne
J.~P., 1995, MNRAS, 277, L5

\bibitem[\protect\citeauthoryear{Revnivtsev, Gilfanov, \&
Churazov}{2000}]{2000A&A...363.1013R} Revnivtsev M., Gilfanov M.,
Churazov E., 2000, A\&A, 363, 1013

\bibitem[\protect\citeauthoryear{Strohmayer}{2001}]{2001ApJ...552L..49S}
Strohmayer T.~E., 2001, ApJ, 552, L49

\bibitem[\protect\citeauthoryear{Sunyaev \&
Revnivtsev}{2000}]{2000A&A...358..617S} Sunyaev R., Revnivtsev M.,
2000, A\&A, 358, 617

\bibitem[\protect\citeauthoryear{Vaughan et
al.}{2005}]{2005MNRAS.356..524V} Vaughan S., Iwasawa K., Fabian
A.~C., Hayashida K., 2005, MNRAS, 356, 524

\bibitem[\protect\citeauthoryear{Wilson \&
Done}{2001}]{2001MNRAS.325..167W} Wilson C.~D., Done C., 2001,
MNRAS, 325, 167

\bibitem[\protect\citeauthoryear{Zi{\'o}{\l}kowski}{2005}]{2005MNRAS.358..851Z}
Zi{\'o}{\l}kowski J., 2005, MNRAS, 358, 851

\bibitem[\protect\citeauthoryear{{\.Z}ycki \&
Nied{\'z}wiecki}{2005}]{2005MNRAS.359..308Z} {\.Z}ycki P.~T.,
Nied{\'z}wiecki A., 2005, MNRAS, 359, 308




\end{thebibliography}
\end{document}